\documentclass[conference]{IEEEtran}
\IEEEoverridecommandlockouts
\usepackage{cite}
\usepackage{amsmath,amssymb,amsfonts}
\usepackage{algorithmic}
\usepackage{graphicx}
\usepackage{textcomp}
\usepackage{multicol}  
\usepackage{subfigure}
\usepackage{marvosym}
\usepackage{multirow} 
\usepackage{CJK}
\usepackage{booktabs}
\usepackage{float}
\usepackage{xcolor}
\usepackage{amsmath}
\usepackage{caption2}
\usepackage{threeparttable}
\def\BibTeX{{\rm B\kern-.05em{\sc i\kern-.025em b}\kern-.08em
    T\kern-.1667em\lower.7ex\hbox{E}\kern-.125emX}}
\begin{document}

\title{Fake Reviewer Group Detection in Online Review Systems}

\author{\IEEEauthorblockN{1\textsuperscript{st} Chen Cao}
\IEEEauthorblockA{\textit{Dalian University of Technology} \\
Dalian, China \\
caochen330@outlook.com}
\and

\IEEEauthorblockN{2\textsuperscript{nd} Shihao Li}
\IEEEauthorblockA{\textit{Dalian University of Technology} \\
Dalian, China \\
shihao\_leee@outlook.com}
\and
\IEEEauthorblockN{3\textsuperscript{rd} Shuo Yu\textsuperscript{\Letter}}
\IEEEauthorblockA{\textit{Dalian University of Technology}\\
Dalian, China \\
shuo.yu@ieee.org}
\and
\IEEEauthorblockN{4\textsuperscript{th} Zhikui Chen}
\IEEEauthorblockA{\textit{Dalian University of Technology}\\
Dalian, China \\
zkchen@dlut.edu.cn
}}

\maketitle

\begin{abstract}
Online review systems are important components in influencing customers’ purchase decisions. To manipulate a product's reputation, many stores hire large numbers of people to produce fake reviews to mislead customers. Previous methods tackle this problem by detecting malicious individuals, ignoring the fact that the spam activities are often formed in groups, where individuals work collectively to write fake reviews. Fake reviewer group detection, however, is more challenging due to the difficulties in capturing the underlying relationships in groups. In this work, we present an unsupervised and end-to-end approach for fake reviewer group detection in online reviews. Specifically, our method can be summarized into two procedures. First, cohensive groups are detected with modularity-based graph convolutional networks. Then the suspiciousness of each group is measured by several anomaly indicators from both individual and group levels. The fake reviewer groups can be finally detected through suspiciousness. Extensive experiments are conducted on real-world datasets, and the results show that our proposed method is effective in detecting fake reviewer groups compared with the state-of-the-art baselines. 
\end{abstract}


\section{Introduction}
Recent decades have witnessed rapidly increasing popularity in online review systems. More and more people write reviews and refer to reviews on websites such as Yelp, Amazon before purchasing since they give customers useful information and first-hand experiences about goods. Therefore, online reviews weigh heavily in promoting sales. However, this financial importance also lures some stores and organizations to hire people to write 
quantities of fake reviews. This activity is commonly called review spam. By indirectly promotes or demotes a product's reputation, review spam effectively affects revenues in business. 

This malicious intrusion into online review systems has attracted much attention from researchers. The problem of detecting such users was first formulated by Jindal et al.\cite{jindal2008}, and then various solutions have been proposed\cite{li2011}\cite{feng2012}\cite{ako2013}. Traditional methods leverage handcrafted individual features to classify reviews or reviewers. The features commonly used can be sorted into three categories: behavior-based features\cite{li2011}\cite{muk2013}\cite{xie2012} that characterize reviewers' behavior, language-based\cite{feng2012}\cite{www20} features that utilize the linguistic patterns of review content, and relation-based\cite{ako2013} features that capture the underlying relationships among users via graphs. 

Most previous studies concentrate on detecting individual fake reviewers, but in real world, fake reviewers often participate in spam activities together as a group. And fake reviewer groups can do much more harm to online review systems comparing to individual ones. However, it is non-trivial to detect fake reviewer groups for two reasons. First, a user can appear normal if analyzed at the individual level, but his or her suspiciousness may only be evident when viewing collectively with some other users. Second, target products and participants of spam activities often overlap among different fake reviewer groups, which makes it difficult to accurately discovering such members. Previous approaches in faker reviewer group detection \cite{wang2016} have proposed several group suspiciousness determinants that capture group-level features based on graphs. However, they do not harness group-level and individual-level features together, leaving out many valuable footprints from metadata. Also, the sizes of detected groups in these methods are mostly quite small \cite{wang2018}. Meanwhile, studies have shown that the rising of crowdsourcing platforms accelerates organizing large groups to conduct powerful malicious web activities \cite{dirty}, including review spam \cite{serf}\cite{crd}, while small reviewer groups can hardly have significant impacts on review systems by contrast. Therefore, large fake reviewer groups are more common in reality, not to mention that detected groups with small sizes have a higher possibility to be generated purely by accident.  

The aforementioned graph-based models all exploit conventional graph approaches. With progress in Graph Neural Networks (GNNs), many researchers apply GNNs to the review spam problem. Unlike previous approaches, GNN-based methods can aggregate messages from neighbors to represent nodes. Most existing works in review spam leverage GNNs for node representation and classification, and this neglects its capability in graph pooling and clustering. Meanwhile, studies have shown that GNNs can utilize higher-order structural information generated by clusters \cite{decoupling}.

In this paper, we propose a fake reviewer group detection method named REAL (modulaRity basEd grAph cLustering for fake reviewer group detection). In detail, REAL first employs GNNs and spectral modularity for graph clustering to find candidate groups. Next, it measures each group's suspiciousness by considering both the group-level indicators and the average values of individual-level indicators collectively via anomaly scores. Based on the ranking of anomaly scores,  REAL can extract the most suspicious fake reviewer groups. Specifically, we refer to modularity metrics to discover overlapping clusters, which can better characterize real-world review spam scenarios. Also, by setting the number of clusters properly, we can extract groups with large sizes. Furthermore, we elaborate a group-level anomaly indicator Group Anomaly Compactness that characterizes the closeness in the relation among group members. REAL is comprehensively effective at detecting large fake reviewer groups with high precision. 

\begin{figure*}[htbp]
	\centering  
	\includegraphics[scale=0.4]{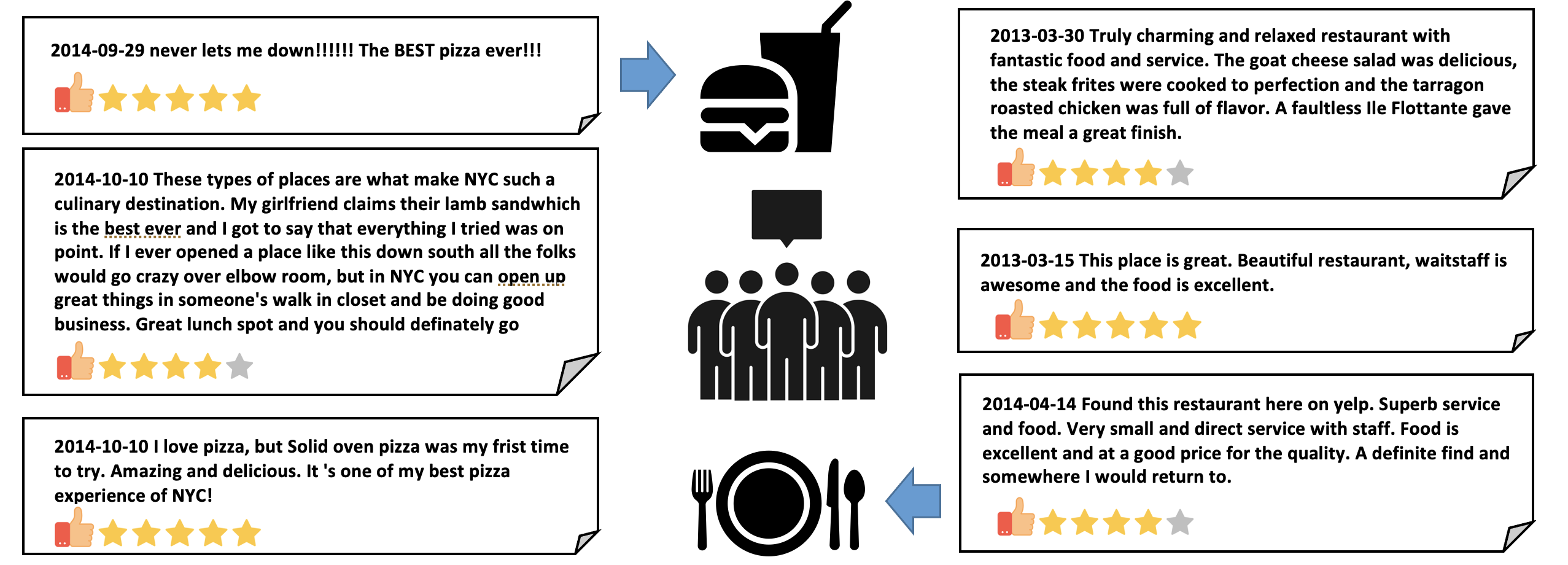} 
	\caption{An example of a fake reviewer group's  spam activity in two restaurants.}
	\label{fig-res3}
\end{figure*}

In summary, our contributions are three-fold:

\begin{itemize}
	\item We present REAL, an unsupervised and end-to-end fake reviewer group detection model leveraging modularity-based graph clustering. REAL introduces spectral modularity for overlapping clustering to facilitate detecting candidate fake reviewer groups.
	\item We unify both group-level and individual-level anomaly indicators together to compute an anomaly score for each candidate group. These indicators harness clues from users' behaviors, temporal patterns, and latent relational closeness to evaluate the suspiciousness of a group.
	\item We conduct extensive experiments on three real-world datasets, namely, YelpNYC, YelpZip, and YelpCHI. The results show that REAL is effective at detecting fake reviewer groups and outperforms three baselines by striking a balance between precision and group size. 
\end{itemize}

The rest of this paper is organized as follows. Section \MakeUppercase{\romannumeral2} goes through the related work in review spam detection. In Section \MakeUppercase{\romannumeral3}, we first introduce some preliminaries of our model REAL, and then demonstrate its framework. Section \MakeUppercase{\romannumeral4} discribes our experiments and results. The conclusion of our work is in in Section \MakeUppercase{\romannumeral5}.

\section{Related Work}
\subsection{Fake Reviewer Detection}
Since Jindal and Liu\cite{jindal2008} propose the fake review detection problem, fake reviews in online shopping websites have gained increasing 
concern. Most previous studies in this area can be categorized into the following types: behavior-based, linguistics-based, and relation-based. For behavior-based methods, Li et al.\cite{li2011} consider semi-supervised models that utilize reviews' and users' views by co-training processes. Xie et al.\cite{xie2012} study the temporal pattern of users' behavior to uncover fake reviewers. For linguistics-based methods, Feng et al.\cite{feng2012} exploit features generated from Context Free Grammar parse trees to make improvements. Neiterman et al.\cite{www20} develop a multilingual speech-based detection model. Considering graphs' advantages in capturing inter-dependent properties from metadata, some proposed graph-based approaches. Wang et al.\cite{wang2011} present a heterogeneous graph model and use an iterative approach to discover the propagation in users.  Li et al.\cite{li2014} build a user-IP-review graph that connects reviews from the same users and IPs. 

Most existing methods focus on fake reviewer detection at the individual level rather than the group level. However, most review spam activities are well-organized and thus fake reviewers form collusive groups, especially along with the rising of crowdsourcing/crowdturfing systems \cite{serf}. Fake reviewer groups are more powerful in distorting a product's reputation. They are also much more difficult to discover as they may appear benign at the individual level. Mukherjee et al.\cite{muk2012} propose a frequent itemset mining method to detects fake reviewer groups. Wang et al.\cite{wang2016} generate loose fake reviewer groups from graphs in a divide and conquer manner. Wang et al.\cite{wang2018} later introduce a new approach that decompose the entire reviewer graph into small fake reviewer groups by a minimum cut algorithm. Dhawan et al.\cite{dhawan2019} propose DeFrauder that harnesses group indicators from behavior and graph features to discover and rank fake reviewer groups.  However, they do not consider group-level and individual-level features in a unified manner, ignoring the effectiveness of traditional features in group detection.

\subsection{Graph Neural Networks}
With the resurgence in deep learning, researchers have begun to apply GNNs to graph-based anomaly detection\cite{query}. Compared with traditional approaches, GNNs can effectively aggregate information from neighbors and extract features in graphs. Basic GNNs \cite{gcn} complete aggregation via the mean function and degree of neighbors, after which many improved algorithms followed. GraphSAGE \cite{graphsage}
merges messages of self-node features from previous layers and introduces the
concept of sampling in GNNs to alleviate the cost problem in computing.  GATs \cite{gat} use a pair-wise function on nodes that updates weights to learn node features. 

Besides the application in fake news detection\cite{fakenews}, financial assessment\cite{alike}, many researchers also use GNNs for review spam problem. For example, Wang et al.\cite{fdgars} trains a GCN\cite{gcn} model to find fraudsters. Li et al.\cite{gas} constructs two graphs and uses GCN to learn reviews' local and global context. Dou et al. \cite{enhancing} strengthens the aggregation process in GNN to defend from camouflages. However, GNNs in fake review detection are mostly used for node representation and individual reviews/reviewers classification. This ignores the problem of fake reviewer groups and fails to recognize GNNs' effectiveness in graph pooling especially graph clustering\cite{offer}. Moreover, Wang et al.\cite{decoupling} demonstrates that jointly performing representation and classification with GNNs would deteriorate training effectiveness if there is users' behavior pattern and label semantics are not conformed, which is a common case in real-world settings.

 Compared with previous models, REAL contributes comprehensively to the issues above. It uses GNNs as a building block and utilizes information from users' behavior and graph structure collectively to discover the most suspicious groups.

\section{Preliminaries}
\subsection{Graph Construction}
 
We construct an attributed "product-rating" graph $\mathcal{G}=\mathcal{(V,E)}$ that aims to incorporate more information from metadata. As demonstrated in Fig. 2, a node $v_{ij}\in \mathbf{V}$ represents a product $p_i$ and one kind of its ratings $r_j$, namely, a product-rating pair $(p_i,r_j)$. The node attribute $\mathbf{a}^n_{ij}$ is defined as the reviewer set where each member has given product $p_i$ the rating $r_j$. The edge $e_{(ij,mn)}$ indicates reviewer sets in which members share identical rating behavior on the two products $p_i, p_m$, and the edge attribute $\mathbf{a^e_{(ij,mn)}}$ denotes the co-review and co-rating reviewer set $\mathcal{R}_{(ij,mn)}$. For example, if reviewer Alice and reviewer Bob both gave $p_i$ the score of $r_j$ and $p_m$ the score of $r_n$, then Alice and Bob should be included in the $\mathcal{R}_{(ij,mn)}$. Note that we assume that each user can only give a product one rating.
 
 \begin{figure}[htbp]
 	\centering
 	\includegraphics[scale=0.38]{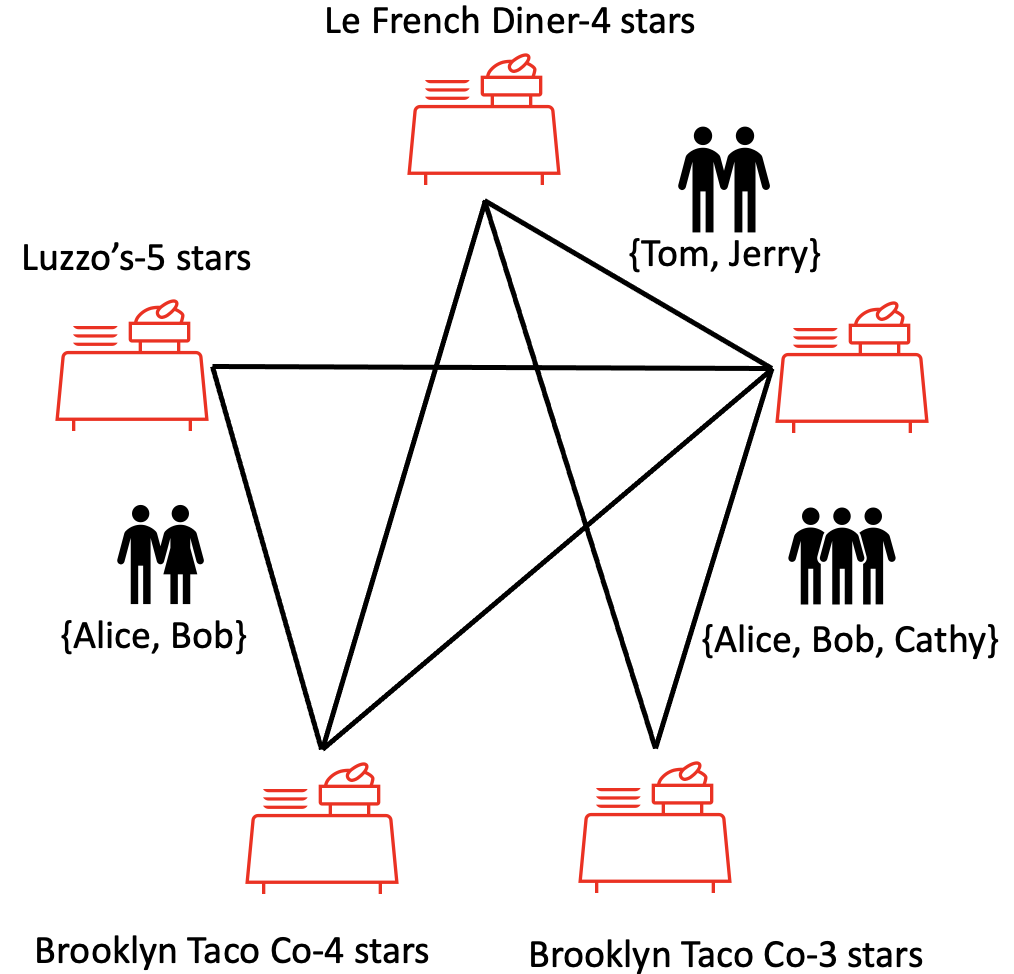}
 	\caption{An illustration of the constructed graph based on Yelp datasets.}
 	\label{fig:label}
 \end{figure}

\subsection{Spectral Modularity for Clustering}
We first extract candidate groups based on cluster results before using group-level anomaly indicators. We handle this task by using methods in graph clustering, which is also commonly known as dividing networks into multiple communities. A community in networks is popularly defined as a group that has tighter relation within, which aligns with our description of fake reviewer groups. 

Traditional approaches in graph clustering use cut-based metrics to detect communities in graphs\cite{2000normalized}. However, the good cuts these methods assume\cite{jure2008} are vulnerable as one node may belong to more than one community. In our scenario, a product could be involved in several spam activities and a fake reviewer could participate in several collusive groups. Meanwhile, modularity\cite{newman2006} based metrics tackle this problem by maximizing intra-cluster relation and minimizing inter-cluster relation. Specifically, given each node $i$ its degree $d_i$, a random graph with $m$ edges is generated, and the node pair($u,v)$ is linked with probability $d_ud_v/2m$. Modularity is then calculated as follows:
\begin{equation}
Q=\frac{1}{2m}\sum_{ij}[A_{ij}-\frac{d_id_j}{2m}\delta(c_i,c_j)]
\end{equation}

\begin{equation}
\delta(u,v)=
\begin{cases}
	1& u = v\\
	0& \text{otherwise}
\end{cases}
\end{equation}
where $c_x$ denotes the cluster a node is assigned to. Modularity calculates the divergence of the edges within a cluster from the expected one, and
we harness it to characterize and detect overlapping collusive groups in $\mathcal{G}$. 

To solve the NP-hard problem of maximizing the modularity, let $d$ be the degree vector, modularity matrix $\mathbf{C}\in 0,1^{n\times k}$ be the cluster assignment and $\mathbf{B}=\mathbf{A}-\frac{\mathbf{d}\mathbf{d}^\top}{2m}$, and modularity can be refined as
\begin{equation}
\mathbf{Q}=\frac{1}{2m}\text{Tr}(\mathbf{C}^\top \mathbf{B}\mathbf{C})
\end{equation}
where Tr(*) denotes the trace of matrix *.

While this process utilizes completely the structure of graphs, we still face the cost problem when adapting it to our attributed graphs.

\section{Methodology}
In this part, we dilate on the framework of REAL. It first extracts candidate groups on the constructed graph by graph clustering with deep modularity networks. Then REAL measures the suspiciousness of candidate groups by computing anomaly scores combining group-level indicator Group Anomaly Compactness and effective individual-level indicators. After ranking the anomaly score of all candidate groups, the most suspicious groups are detected. The overall framework is illustrated in Fig.3.

\begin{figure*}[htbp]
	\centering  
	\includegraphics[scale=0.47]{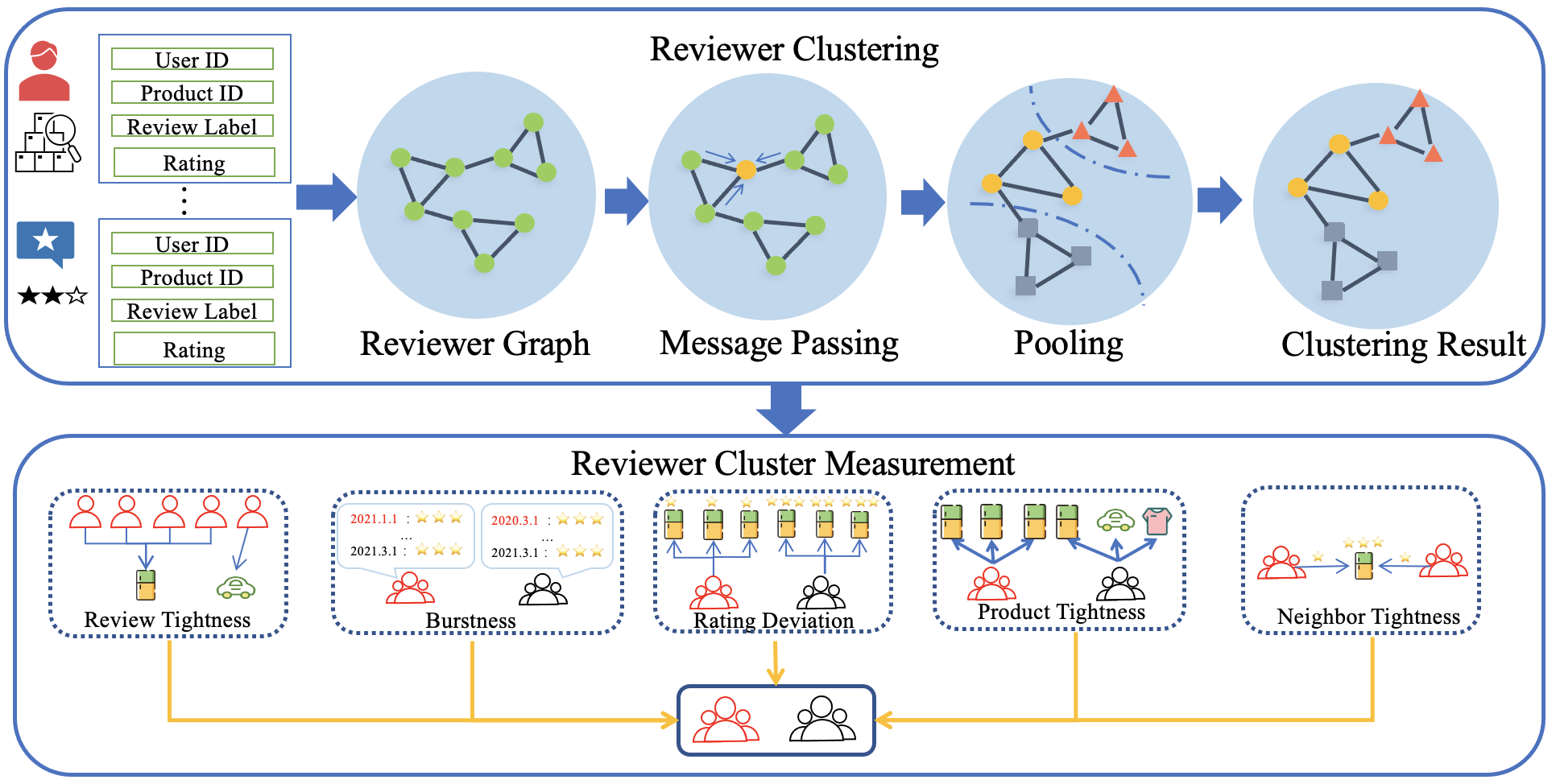} 
	\caption{REAL's framework. With metadata including information about users, products, review labels, ratings and dates, a "product-rating" graph is constructed. REAL then performs graph clustering with GNNs for clustering. Based on the cluster results, it measures candidate fake reviewer groups with indicators both at group levels and individual levels. Finally, REAL merges outcomes of all the indicators with an elaborated scoring function that gives more weight to group indicators and ranks the candidate groups. }\label{fig-res3}
\end{figure*}

\subsection{Graph Convolutional Networks}
The aim of generalizing convolutions to graph domain is to encode the
nodes with signals in the receptive fields. Given a graph $\mathcal{G}=(\mathcal{V,\mathcal{E}})$ with nodes count $N=|\mathcal{V}|$, where each node is represented with a feature vector. This way the convolution operator can be represented by Hadamard product in the Fourier domain.

For a GCN with $L$ convolutional layers, each layer embeds nodes by aggregating its neighbors from previous layer. Consider the common practice for a GCN:
\begin{equation}
	\mathbf{H}^{(t+1)}= \sigma(\widetilde{\mathbf{A}}\mathbf{H}^{(t)}\mathbf{W}^{(t)})
\end{equation}

where
$\widetilde{\mathbf{A}}=\hat{\mathbf{D}}^{-\frac{1}{2}}\hat{\mathbf{A}}\hat{\mathbf{D}}^{-\frac{1}{2}}$, $\hat{\mathbf{A}}=\mathbf{A}+\mathbf{I}$, $\hat{\mathbf{D}}$ is the diagonal node degree matrix of $\hat{\mathbf{A}}$. $\mathbf{H}^{(0)}=\mathbf{X}$ is the input in the first layer. $\mathbf{X}$ is the feature matrix. $\mathbf{W}^{(t)}$ is a learnable matrix shared among all nodes at layer $t$, and $\sigma(\cdot)$ is the activation function.

In REAL, we use SeLU\cite{sel} for activation, which is proven to benefit convergence. Also, we skip the self-loop and insert a weight matrix $\mathbf{W}_{s}$ for skip connection. 

Therefore, the $t$-th output layer for $\mathbf{H}^{(t+1)}$ is

\begin{equation}
	\mathbf{H}^{(t+1)}=\text{SeLU}(\mathbf{\widetilde{A}}\mathbf{H}^{(t)}\mathbf{W}+\mathbf{H}\mathbf{W}_{s})
\end{equation}

\subsection{Deep Modularity Network}

To conduct graph clustering on $\mathcal{G}$ and select candidate collusive groups, we refer to deep modularity networks that use modularity-based loss function for optimization. In detail, the model gets the cluster assignment matrix via softmax on graph convolution networks. We obtain $\mathbf{C}$ as follows:
\begin{equation}
\mathbf{C}=\text{softmax}(GCN(\mathbf{\widetilde{A}},\mathbf{X}))
\end{equation}
We then introduce spectral modularity in its optimization function. However, allocating every node to the same cluster may trigger the local minima problem which harms optimization\cite{minpool}. Therefore we assign spectral clustering a regularization to guarantee the informativeness in clustering. To avoid being too restrictive when performed with softmax, the collapse regularization is relaxed as$	\frac{\sqrt{k}}{n}\left \| \sum_{i}\mathbf{C}_i^\top \right \|_F-1$, where $\left \|\cdot \right \|_F$ is the Frobenius norm, $n$ is the number of nodes and $k$ is the number of the clusters. 

So the complete loss function is defined as follows:
\begin{equation}
	\mathcal{L}=-\frac{1}{2m}\text{Tr}(\mathbf{C}^\top \mathbf{B}\mathbf{C})+\frac{\sqrt{k}}{n}\left \| \sum_{i}\mathbf{C}_i^\top \right \|_F-1
\end{equation}

\subsection{Anomaly Indicators}
Our next step is to evaluate candidate groups with anomaly indicators and sort out the most suspicious ones. After graph clustering, nodes in the graph $\mathcal{G}$ are assigned to the clusters. We then intersect the reviewer sets of targeted products in the same clusters, which are also the attributes of nodes, and get the group members of each cluster. To grasp messages from reviewers' behavior, content and underlying relation, we adopt the indicators discussed below. Note that according to Xu et al.\cite{chinese2013}, language features can be easily imitated by fake reviewers. Furthermore, many online review systems only allow rating for products without comments in texts. Therefore, in our settings, we focus on rating behaviors and users' temporal characteristics, discarding the language features. 

\subsubsection{Group Anomaly Compactness Indicator}

As members in fake reviewer groups may appear benign when considered separately but become conspicuous outliers in a group context, we attach greater importance to group-level indicators than individual ones. We employ several group anomaly indicators that measure behavioral suspiciousness in group context\cite{wang2016}. After experimenting on previously illustrated group indicators, and we find only a small fraction of them characterizes fake reviewer groups with high accuracy. To efficiently demonstrate fake reviewer groups, we summarize these group indicators and elaborate a group-level indicator that quantifies the closeness within a group, namely, Group Anomaly Compactness. It effectively fuses the information about a group's product set, review set and members' behaviors. Note that the formation of small groups is highly likely to be coincident because it's common for two or three people to co-review a product by chance, but by design for large groups. Therefore, a penalty function is designed as follows:

\begin{equation}
	L(g)=\frac{1}{1+e^{-(|\mathcal{R}(g)+|\mathcal{P}(g)|-3})}
\end{equation}

where $\mathcal{R}(g)$ is the reviewers in group $g$, and $\mathcal{P}(g)$ is the respective products rated by the users in $\mathcal{g}$. $\mathcal{P}_i$ is the i-th product set a user has reviewed. We then measure the group-level suspiciousness via the following indicators and then compute Group Anomaly Compactness. 

\begin{itemize}

	\item \textbf{Review Tightness (RT)} If a large proportion of people jointly review some products, it indicates spam group activities. Given a candidate fake reviewer groups $g$, review tightness calculates its ratio of the total count of reviews to the product of the product set and the user set multiplies $L(g)$:
\begin{equation}
RT(g)=\frac{\sum_{i\in {\mathcal{R}(g)}}|\mathcal{P}_i|}{|\mathcal{R}(g)||\mathcal{P}(g)|}\cdot L(g) 
\end{equation}

	\item \textbf{Product Tightness (PT)} When a group focuses on certain products while reviewing few other ones, it is likely to be a fake reviewer group as members in normal groups would rate a variety of goods and the intersection of the product they review wouldn't be that large. Given a group $g$, its product tightness is the number of products commonly rated by the group $g$ to the total amount of products rated by all the members:
\begin{equation}
	PT(g)=\frac{|\cap_{r\in {\mathcal{R}_g}}\mathcal{P}_i|}{|\cup_{r\in{\mathcal{R}_g}}\mathcal{P}_i|}
\end{equation}

\item \textbf{Neighbor Tightness (NT)} When the products two groups reviewed are similar, both groups are possible fake reviewers because it's rare for two groups of people to share highly similar product sets in normal scenarios. Therefore, we define neighbor tightness as the average of Jaccard Similarity (JS) of the product sets in each user pairs:
\begin{equation}
	NT(g)=\frac{2\sum_{i,j\in {\mathcal{R}(g)}}JS(\mathcal{P}_i,\mathcal{P}_j)}{\mathop{|\mathcal{R}_g|}} L(g)
\end{equation}

Then with the following group-level indicators, we can compute Group Anomaly Compactness $\Pi$ as follows:

\begin{equation}
	\Pi = RT(g)*PT(g)*NT(g)
\end{equation}
\subsubsection{Individual Anomaly Indicators}

Though members in a fake reviewer group may appear normal at the individual level, we still exploit clues from the users' behavioral and temporal features which are complementary to our work. Note that we would give more weight to Group Anomaly Compactness than individual-level indicators when computing the final anomaly scores. 

\item \textbf{Average User Rating Deviation (AVD)} Defining rating deviation $d_{ij}$ as the absolute deviation of $i$'s rating from the product $j$'s average rating\cite{li2011}, average user rating deviation is the average value of the reviews' deviation that belongs to the reviewer.

\item \textbf{Burstness (BST)} Fake reviewers often appear during a short term on the website while benign users are often active for a longer term.
\begin{equation}	
BST(i)=\left\{
\begin{array}{lcl}
	0& &if\;E(i)-F(i)>\tau, \\
	1-\frac{E(i)-F(i)}{\tau}& &otherwise\\
\end{array}
\right.
\end{equation}
where $F(i)$ is the date of user $i$'s first review and $E(i)$ is the date of $i$'s last review. $\tau$ is the threshold. In our work, we set it to 30 days.
\end{itemize}

\subsection{Candidate Groups Ranking}

Given a candidate group $g$ and its members $r_i\in\mathcal{R}(g)$, we measure its suspiciousness by ranking the anomaly score $\Omega$. To be specific, after calculating its scores of Group Anomaly Compactness $\Pi$ and the individual-level indicators, each score is scaled between 0 and 1 with min-max Normalization. Then $g$'s anomaly score $\Omega$ is calculated as follows:	

\begin{equation}
\Omega = 3*\Pi+\frac{ \sum_{r_i \in \mathcal{R}(g) }ARD(r_i)}{ | \mathcal{R} (g) | } +\frac{ \sum_{r_i \in \mathcal{R}(g) }BST(r_i)}{ | \mathcal{R} (g) | }
\end{equation}
where Group Anomaly Compactness multiples 3 before added to the final anomaly scores as we emphasize more on group-level anomaly messages.

The higher a group's anomaly score is, the more suspicious it is to be a fake reviewer group.

\section{Experiments}

\subsection{Datasets}
We evaluate our method on three real-world datasets from Yelp.com collected by \cite{kdd2015}. YelpNYC contains reviews for restaurants in New York City. YelpCHI, collected by \cite{2013yelp}, comprises reviews for a set of hotels and restaurants in the Chicago area. YelpZip collects restaurant reviews from plenty of areas ordered by zip code. Details for each datasets is illustrated in Table \MakeUppercase{\romannumeral1}. Note that the labels in the datasets are near-ground-truth since they are generated by Yelp's filtering algorithms.
\begin{table}[ht]
	\centering  
	\begin{tabular}{cccc}
		\toprule  
		Dataset& \#Reviews& \#Reviewers&\#Products\\
		\midrule 
		YelpNYC& 359,052& 160,225&923\\
		YelpCHI& 67,395&38,063&201\\
		 YelpZip& 608,598&260,227& 5,044\\
		\bottomrule 
	\end{tabular}
	\caption{Statistics of three datasets}
	\label{Tab: label}
\end{table}

\subsection{Baeselines}
We compare our methods with three baselines: 
\begin{itemize}
	\item \textbf{GraphStrainer} by Ye et al. \cite{lek2015nfs} first uses a graph-based method to discover target products and then detects fake reviewer groups via hierarchical clustering based on induced subnets. It does not use any handcrafted features.
	\item \textbf{ColluEagle} by Wang et al.\cite{wang2020} is a markov random field based method exploiting group-level behavior features. It uses a density-based clustering approach\cite{scan} to extract candidate groups rather than deep approaches. Note that in our experiments we use vanilla ColluEagle without node prior.
	\item \textbf{DeFrauder} by Dhawan et al.\cite{dhawan2019} first uses group indicators to extract groups, and then performs graph embedding via node2vec\cite{node2vec} to calculate density-based spam scores. DeFrauder exploits only group features, ignoring individual-level ones.
\end{itemize}
\subsection{Evaluation}

With only labels about the truthfulness of each review, it is hard to decide whether several people have worked collaboratively. We consider that the more fake reviewers in a group, the more suspicious the group is. Therefore, we define the precision of results as the ratio of fake reviewers in the group to all group members. In our experiments, we mark the user that has written at least one fake review as fake.
\begin{table*}[htbp]
	\centering  
	
		\begin{tabular}{ccccccc}
			\toprule
			\multirow{2}{*}{Method}&
			\multicolumn{2}{c}{ YelpNYC}&\multicolumn{2}{c}{ YelpZip}&\multicolumn{2}{c}{ YelpCHI}\cr
			\cmidrule(lr){2-3} \cmidrule(lr){4-5}  \cmidrule(lr){6-7}
			&Group Size&Precision&Group Size&Precision&Group Size &Precision\cr
			\midrule
			DeFrauder&\bf{133}&	0.1955	&20&	0.4500&	21	&0.4762\cr
			GroupStrainer&23&	0.5652&	26&0.5769&\bf{24}&0.4583\cr
			ColluEagle&16&0.5625&17&\bf{0.8235}&16&0.625\cr 
			\midrule
			REAL&25&	\bf{0.7600}&	\bf{39}&0.5641&\bf{24}&\bf{0.6667}\cr
			\bottomrule
		\end{tabular}
		\caption{Group size and precision results of the top group in each algorithm. Since ColluEagle fails to discover fake reviewer groups with more than 20 people in all 3 datasets, we choose the largest group in its results for comparison.}  
	\label{tab:label}
\end{table*}
We consider detected groups with small sizes are much more likely to form by coincidence and also have only slight effects most of the time, while large fake reviewer groups are more common and harmful as mentioned before. Meanwhile, according to Dunbar's number, a person's social circle typically consists 15 good friends and 5 best friends. Considering the sum of these two sets of people as a small social clique, we set 20 as the lower bound for a relatively both large and tight fake reviewer group, and results with less than 20 members will not be taken into account. However, many existing group-level detecting methods\cite{wang2016,wang2018} can only uncover groups with less than 10 people, especially 2 to 5 people\cite{wang2018}, and few detected groups have more than 20 members. Hence we compare the precision of the detected groups that ranked first among the outputs with more than 20 members to alleviate these issues in the baseline for better comparison. GroupStrainer is a cluster-based method, so we select the group that has the highest precision among those have more than 20 members.

\subsection{Experimental Results}

For deep modularity networks in the clustering step, we analyze the effect of the number of clusters on YelpNYC and YelpZip dataset since it can significantly affect the group size in final results, which we consider crucial. We set $k$ in collapse regularization to 0.5, dropout to 0.5 and compare the precision of clustering results, skipping measuring indicators. The precision we calculate at this step is the mean precision value of the top 10 clusters, which is also the optimal situation for the next stage's detection. Particularly, we present the cluster precision results on YelpNYC and YelpZip.

\begin{figure}[h]
	\centering 
	\subfigure[YelpNYC]{
		\label{Fig.sub.1}
		\includegraphics[width=0.23\textwidth]{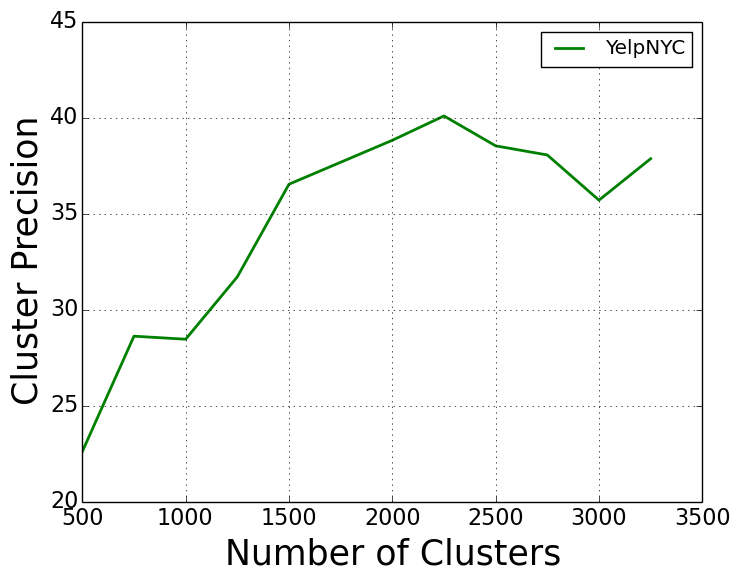}}
	\subfigure[YelpZip]{
		\label{Fig.sub.2}
		\includegraphics[width=0.22\textwidth]{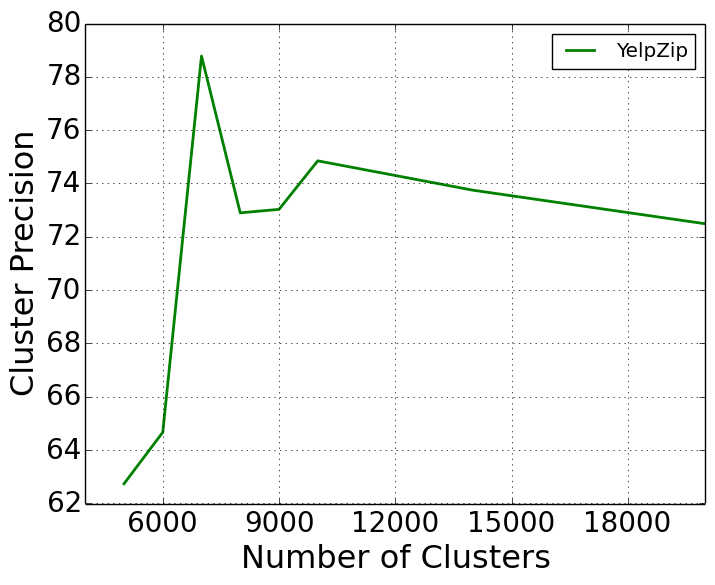}}	\caption{Effect of cluster numbers. The figure above illustrates the precision of cluster results on YelpNYC, and the one below illustrates the precision of cluster results on YelpZip. Specifically, we rank the precision of candidate groups in the results and select the mean value of top 10 groups as cluster precision.  }
	
	\label{Fig.main}
\end{figure}

\begin{table*}[H]
	\centering  
	
		\begin{tabular}{ccccccc}
			\toprule
			\multirow{2}{*}{Method}&
			\multicolumn{2}{c}{ YelpNYC}&\multicolumn{2}{c}{ YelpZip}&\multicolumn{2}{c}{ YelpCHI}\cr
			\cmidrule(lr){2-3} \cmidrule(lr){4-5}  \cmidrule(lr){6-7}
			&Group Size&Precision&Group Size&Precision&Group Size &Precision\cr
			\midrule
			DeFrauder&\bf{133}&	0.1955	&20&	0.4500&	21	&0.4762\cr
			GroupStrainer&23&	0.5652&	26&0.5769&\bf{24}&0.4583\cr
			ColluEagle&16&0.5625&17&\bf{0.8235}&16&\textbf{0.625}\cr 
			\midrule
			REAL&25&	\bf{0.7600}&	\bf{39}&0.5641&\bf{24}&\bf{0.6667}\cr
			\bottomrule
		\end{tabular}
		\caption{Group size and precision results of the top group in each algorithm. Since ColluEagle fails to discover fake reviewer groups with more than 20 people in all 3 datasets, we choose the largest group in its results for comparison.}  
	\label{tab:label}
\end{table*}

As demonstrated in Fig.4, when the cluster number is relatively small, the cluster precision goes up as the number of clusters gets greater. This could be due to that greater cluster numbers lead to smaller group sizes, which is beneficial for theoretical precision while its practical meanings are doubtful. When the cluster number is relatively large, the cluster precision goes down. The reason is that after clustering target product, we intersect the reviewer sets of the products to get candidate groups. Too many clusters will lead to fewer detected groups, and the more clusters we divide, the smaller the group size is. So a decrease in the total number could negatively affect the precision.

We illustrate the distribution of the size of detected groups in Fig.5, This shows that the sizes of most detected groups are among 20 to 150. After ranking the top 100 fake reviewer groups, the sizes gather around 20 to 100 as shown in Fig.6. This could be due to the difficulty in guaranteeing precision when extracting large sizes of groups.
\begin{figure}[htbp]
	\centering
	\includegraphics[scale=0.35]{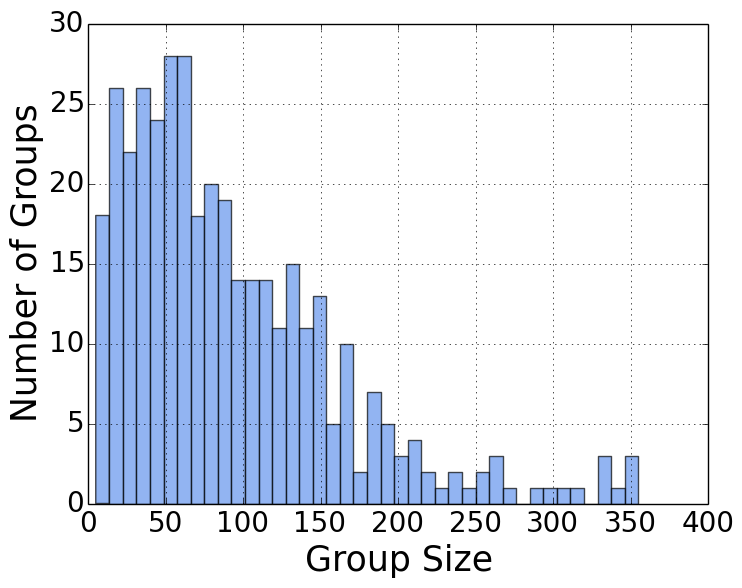}
	\caption{Size distribution of fake reviewer groups in YelpNYC before ranking.}
	\label{fig:label}
\end{figure}

\begin{figure}[htbp]
	
	\centering
	\includegraphics[scale=0.35]{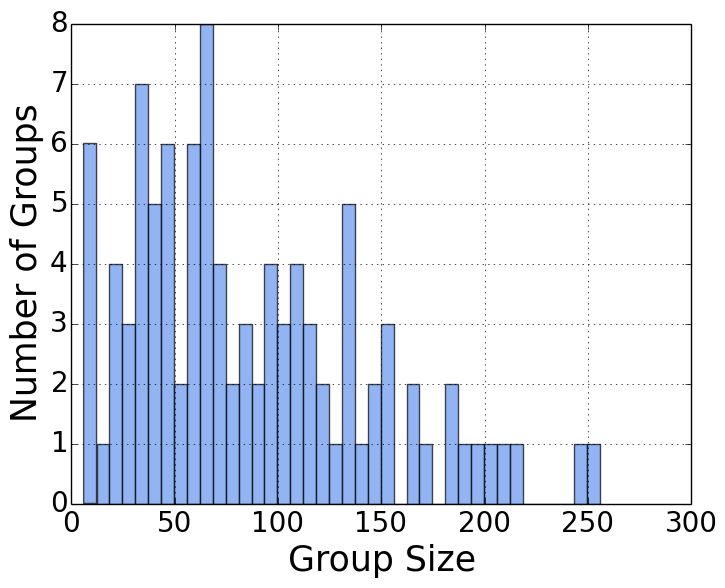}
	\caption{Size distribution of precision top 100 fake reviewer groups in YelpNYC.}
	\label{fig:label}
\end{figure}

Finally, we conduct extensive experiments on YelpNYC, YelpZip, YelpCHI with our baselines, and the precision result of the chosen group as presented in Table \MakeUppercase{\romannumeral2}. The results show that REAL achieves comprehensively good results on all three datasets.

Though ColluEagle achieves high precision in the results, it does not extract any group of large sizes and most groups it detects involve only 2 to 4 people. We cast doubt on the effectiveness of such results because in practice it is a rare case for a fake reviewer group to involve 2 to 4 people and groups with small sizes may be generated by chance. DeFrauder is able to detect many groups with large sizes, but its precision is relatively low compared with REAL. GroupStrainer performs comprehensively, but is still inferior to REAL. Meanwhile, we also notice that REAL's precision is significantly lower in YelpZip compared with ColluEagle, we analyze that since YelpZip is especially larger than the other two, it could be that GCN does not use the same local filter to scan every node and the weights in the filters are the same for all neighboring nodes in the receptive field. As a result, REAL's performance in large graphs is hindered.

Compared with baselines, REAL makes a satisfying balance between the group size and precision.

\section{Conclusion}
Online review systems are susceptible to review spam, and such activities at the group level can cause detrimental effects because large fake reviewer groups are capable of manipulating the products' overall reputation. So uncovering such groups is crucial. In real-world settings, overlapping of target products and members in different groups is a common scenario for spam activities. In this work, we demonstrate REAL, a fake reviewer detection approach via modularity-based graph clustering. REAL is an unsupervised model that can be trained end-to-end. It introduces the concept of spectral modularity into GNNs and performs graph clustering to find out candidate groups. REAL then measures the suspiciousness of each group by unifying group-level and individual-level indicators collaboratively. We validate the effectiveness of our method and compare the experimental results with three baselines, and REAL outperforms all of them by discovering the most suspicious and relatively large group. In the future, we plan to improve the precision on larger sizes of detected groups, and also reduce the computational cost in the procedures.  

\newpage
\bibliographystyle{IEEEtran}
\bibliography{spammers}

\end{document}